\documentclass[prb,a4paper,twocolumn,superscriptaddress]{revtex4}
\usepackage{amsmath} 
\usepackage{amsfonts} 
\usepackage{amssymb} 
\usepackage{bm} 
\usepackage{graphicx}
\usepackage[pagebackref,colorlinks]{hyperref}
\usepackage{epstopdf}
\begin{document}
\title{Vibronic spectroscopy of an artificial molecule}
\author{David Gunnarsson}
\affiliation{Low Temperature Laboratory, Helsinki University of Technology,
FI-02015 HUT, Finland} 
\author{Jani Tuorila}
\affiliation{Department of Physical Sciences, University of Oulu,  FI-90014, Finland} 
\author{Antti Paila}
\affiliation{Low Temperature Laboratory, Helsinki University of Technology,
FI-02015 HUT, Finland} 
\author{Jayanta Sarkar}
\affiliation{Low Temperature Laboratory, Helsinki University of Technology,
FI-02015 HUT, Finland} 
\author{Erkki Thuneberg}
\affiliation{Department of Physical Sciences, University of Oulu,  FI-90014, Finland} 
\author{Yuriy Makhlin}
\affiliation{Landau Institute for Theoretical Physics, Kosygin st. 2,
119334 Moscow, Russia} 
\author{Pertti Hakonen}
\affiliation{Low Temperature Laboratory, Helsinki University of Technology,
FI-02015 HUT, Finland} 
\date{\today}
%
%
\maketitle


{\bf With advanced fabrication techniques it is possible to make nanoscale electronic structures that have discrete energy levels. Such structures are called artificial atoms because of analogy with true atoms. Examples of such atoms are quantum dots in semiconductor hetero\-structures\cite{Kastner93,Ashoori96} and Josephson-junction qubits\cite{Nakamura97,Flees97,Bouchiat98,Nakamura99,vanderWal00}. It is also possible to have artificial atoms interacting with each other. This is an artificial molecule in the sense that the electronic states are analogous to the ones in a molecule\cite{Oosterkamp98}. In this letter we present a different type of  artificial molecule that, in addition to electronic states, also includes the analog of nuclear vibrations in a diatomic molecule. Some of the earlier experiments could be interpreted using
this analogy, including qubits coupled to oscillators\cite{Ilichev03,Chiorescu04,Wallraff04,Lupascu04,Duty05,Sillanpaa05,Johansson06J,Schuster07,Wallraff07} and qubits driven by an intense field\cite{Nakamura01,Saito04,Oliver05,Sillanpaa06,Wilson07}. In our case the electronic states of the molecule are represented by a Josephson-junction qubit, and the nuclear separation corresponds to the magnetic flux in a loop containing the qubit and an LC oscillator. 
We probe the vibronic transitions, where both the electronic and vibrational states change simultaneously, and find that they are analogous to true molecules.
The vibronic transitions could be used for sideband cooling of the oscillator, and we see damping up to sidebands of order 10.}





Consider two different electronic states of a diatomic molecule. For both states one can form a potential energy $U(r)$ as a function of the separation $r$ of the two nuclei. In a bound state they may be approximated by a Morse potential, which is close to a harmonic oscillator potential near the equilibrium separation, but at larger distances the potential is less steep and levels off to allow for dissociation of the molecule. Because of  coupling to the electromagnetic field, there can be transitions between the electronic levels. Associated with this, there is also often a change in the vibrational state of the molecule, and  such transitions are called vibronic. The vibronic transitions are preferred if the two electronic states correspond to different equilibrium separations of the nuclei. The intensity of different transitions is governed by Franck-Condon principle\cite{Franck26,Condon26}. In its classical form the principle says that transitions are possible 
between vibrational states whose  trajectories intersect in phase space and they are
most intense between states having coincident turning points.

\begin{figure}
\begin{center}
\includegraphics[width=1.0\linewidth]{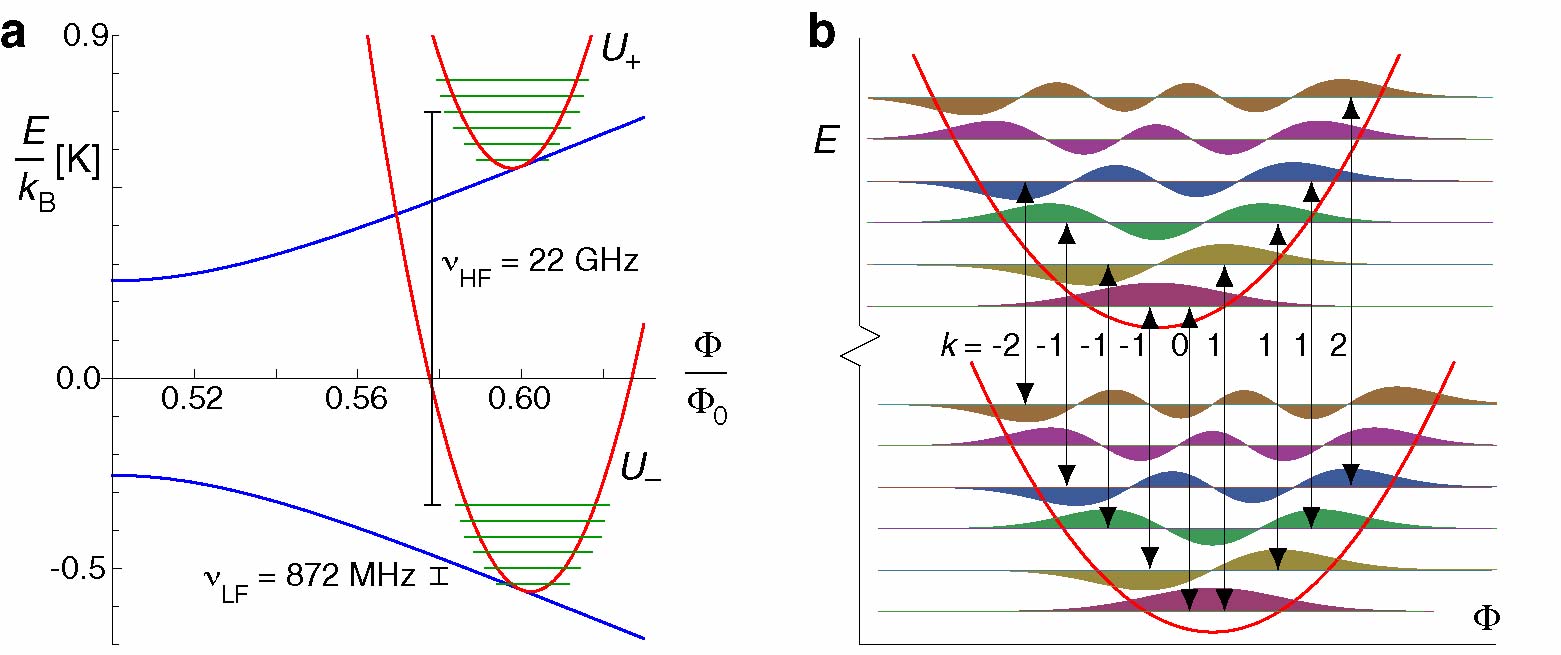}
\caption{The potential energy of the artificial molecule, equation (\ref{e.Vscoweiow}). 
{\bf a}, The total potentials (red) are formed as a sum of qubit energy (blue) and a harmonic  vibrational potential. The six lowest energy eigenstates are shown for both electronic states. The bars represent the energy differences $\Delta E=h\nu$ corresponding to the drives of low and high frequency, 
$\nu_{\rm LF}$ and $\nu_{\rm HF}$, respectively. The parameters are for the bias point $(\Phi_b,n_g)= (0.60\Phi_0,1.16)$ marked by a cross in Fig.\ \ref{MeasPower}a. 
{\bf b}, A close-up at the lowest eigenstates together with wave functions (shaded). The strongest transitions according to the Franck-Condon principle are shown by arrows, and are labeled by the change of vibrational quanta $k$.} \label{f.vibronicpot}
\end{center}
\end{figure}

The corresponding potentials  
in our artificial molecule are shown in Fig.\ \ref{f.vibronicpot}. There are two different total potentials, shown by red lines in Fig.\ \ref{f.vibronicpot}a, corresponding to two different electronic states. In general, the potentials have different force constants and their minima are shifted. The potentials are also anharmonic at higher energies. The only qualitative difference to real atomic potentials is that in our artificial molecule the leveling off of the potential, and the subsequent dissociation, is not achieved at the parameters we depict.  Fig.\ \ref{f.vibronicpot}b shows a closer look at the potential minima, together with the lowest energy eigenstates and the eigenfunctions. The strongest transitions according to the  Franck-Condon principle are shown by arrows.

We have made spectroscopic studies of the artificial molecule. By a low frequency drive we excite the 
vibrations and  by a simultaneous high frequency drive we can induce  the vibronic transitions. By monitoring the reflection of the low frequency wave we can resolve the vibronic levels. 

\begin{figure}
\begin{center}
\includegraphics[width=1.0\linewidth]{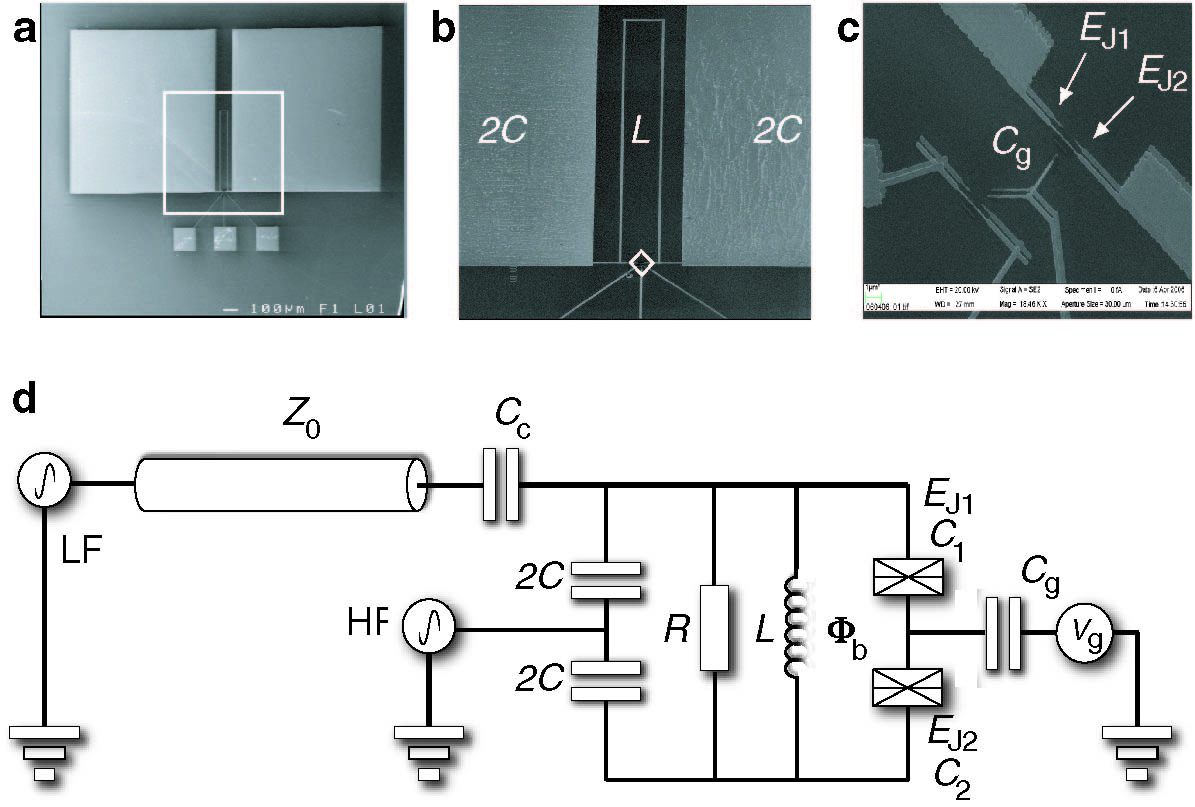}
\caption{SEM micrographs and circuit diagram of the  artificial molecule. 
{\bf a}, General view showing top electrodes of the the oscillator capacitors (2 large pads) with capacitances $2C$ each.
{\bf b}, Enlargement of the white square area in {\bf a} showing the 300 $\mu$m-long microstrip loop that forms the
inductance $L$ of the oscillator.
{\bf c}, Magnification of the white diamond in {\bf b}  showing 
the single Cooper pair transistor (SCPT) formed by two Josephson junctions with coupling energies
$E_{J1}$ and $E_{J2}$ and the gate electrode. 
Also visible is a reference SCPT (lower left) which enables to estimate the
resistances of the junctions in the sample. 
{\bf d}, The circuit diagram shows the SCPT and the LC oscillator connected in parallel.
The resistance $R = 2.8$ k$\Omega$
represents the AC-dissipation of the capacitors as well as of other
parts of the circuit. The artificial molecule is biased by a gate voltage $V_g$ and an external magnetic flux $\Phi_b$. 
For measurement it is connected to a low frequency drive (LF) via a $Z_0=50\
\Omega$ transmission line and a coupling capacitor
$C_c = 7 $ pF. The high frequency drive (HF) is used to excite the vibronic transitions.  } \label{LSET_circ}
\end{center}
\end{figure}

The circuit of our artificial molecule is shown in Fig.\ \ref{LSET_circ}.
The heart of it is a single-Cooper-pair transistor (SCPT). It  consists of a 2 $\mu$m long Al island surrounded by two Josephson junctions with
coupling energies $E_{J1}\approx E_{J2}$,  as well as capacitances
$C_1$ and $C_2$, respectively. The SCPT is fabricated using standard procedure of two angle evaporation\cite{Fulton87}.
The SCPT is coupled in parallel to an LC oscillator. The inductor of the oscillator is formed by
 a superconducting on-chip microstrip loop of length
 $300 ~\mathrm{\mu m}$ giving $L= 169$ pH. The capacitance $C=196.2$ pF is formed by
two parallel plate capacitors  acting
in series. These are made by patterning two
$1.3 ~\mathrm{mm}\times 1.3 ~\mathrm{mm}$ Al plates on the Si$_3$N$_4$/Nb/Si
sandwich substrate. The bottom plate of the capacitor is formed by a
$150 ~\mathrm{nm}$ thick niobium ground plane layer under the
dielectric $\mathrm{Si}_3\mathrm{N}_4$ layer, whose thickness is
around $300 ~\mathrm{nm}$.

Once fabricated, the artificial molecule has still two parameters that can be adjusted in the experiment.
The dc gate voltage $V_g$ changes the gate charge $en_g=C_gV_g$, which controls the tunneling of Cooper pairs to and from the island. The other parameter is the magnetic flux bias $\Phi_b$ through the inductor loop caused by an external magnetic field. This controls the Josephson coupling $E_J$ of the island as the flux shifts the relative phases of the two junctions and leads to complete suppresion at $\Phi_b=\Phi_0/2$ in a symmetric SCPT. Here $\Phi_0=h/2e$ is the flux quantum.
The properties of the SCPT can be calculated in detail using Mathieu functions\cite{Averin85} and including the asymmetry of the junctions. Instead of the general analysis we present here a simplified treatment that is still sufficient to understand the main properties of the circuit. 
In the charge limit $E_J\ll E_c=e^2/(C_1+C_2+C_g)$, the SCPT forms a qubit where only two charge states are of importance. They correspond to having one Cooper pair more or less on the island. In this basis the Hamiltonian of the SCPT  is
$H_{\rm q}=
\frac{1}{2}[\sigma_zE_{\rm el}-\sigma_xE_{J0}\cos(\pi\Phi/\Phi_0)]$, where $E_{J0}=E_{J1}+E_{J2}$, 
$E_{\rm el} =4E_c(1-n_g)$ and $\sigma_i$ are Pauli matrices.
Diagonalizing the Hamiltonian gives two eigenvalues $E^{\rm q}_\pm=\pm
\frac{1}{2}\sqrt{E_{\rm el}^2+E_{J0}^2\cos^2(\pi\Phi/\Phi_0)}$, which correspond to two  "electronic" states of  the artificial molecule. The LC circuit forms a harmonic oscillator, whose Hamiltonian $H_v=q^2/2C+(\Phi-\Phi_b)^2/2L$. Taking the flux $\Phi$ as a generalized coordinate, the charge $q$ on the capacitor is the corresponding canonical momentum. Thus the total potential energy  $U$ of the "molecule" is 
\begin{eqnarray}
U_\pm(\Phi)=\pm
\frac{1}{2}\sqrt{E_{\rm el}^2+E_{J0}^2\cos^2\frac{\pi\Phi}{\Phi_0}}
+\frac{(\Phi-\Phi_b)^2}{2L}.
\label{e.Vscoweiow}\end{eqnarray}
This is plotted in Fig.\ \ref{f.vibronicpot} using parameters appropriate for our circuit. We see that the electronic and vibrational degrees of freedom are coupled as the electronic potential also depends on the flux $\Phi$. The LC oscillator potential vanishes at the bias point $\Phi_b$ but because of  opposite slopes of the qubit energies, the minima of the total potentials are shifted in opposite directions\cite{Nakamura01,Wilson07}. Because of the different curvatures of the  qubit potentials, the vibrational resonance frequencies $\nu_{\pm}$ in the two electronic states are different\cite{Wallraff04,Duty05,Sillanpaa05}.

For measurement the molecule is connected to a transmission line. 
The measurement is done by sending in a low frequency signal typically at the frequency $\nu_{\rm LF}=872$ MHz.
This is slightly below the LC oscillator resonance frequency $\nu_0=874$ MHz.
The reflected signal was continuously monitored using
 a network analyzer set to detect magnitude and phase in a band width of
10 kHz.  Being close to resonance, the reflected signal depends essentially on the vibrational frequency determined by the curvature of the qubit energy. As this varies as a function of the bias point $(\Phi_b,n_g)$, one can use the data on the electronic ground state to fit the parameters of the qubit. We find  $E_c/k_B = 0.8$ K, $E_{J0}/k_B= 3.2$
K, and the
asymmetry parameter $d=(E_{J1}-E_{J2})/(E_{J1}+E_{J2})=0.04$.

\begin{figure}
\begin{center}
\includegraphics[width=1.0\linewidth]{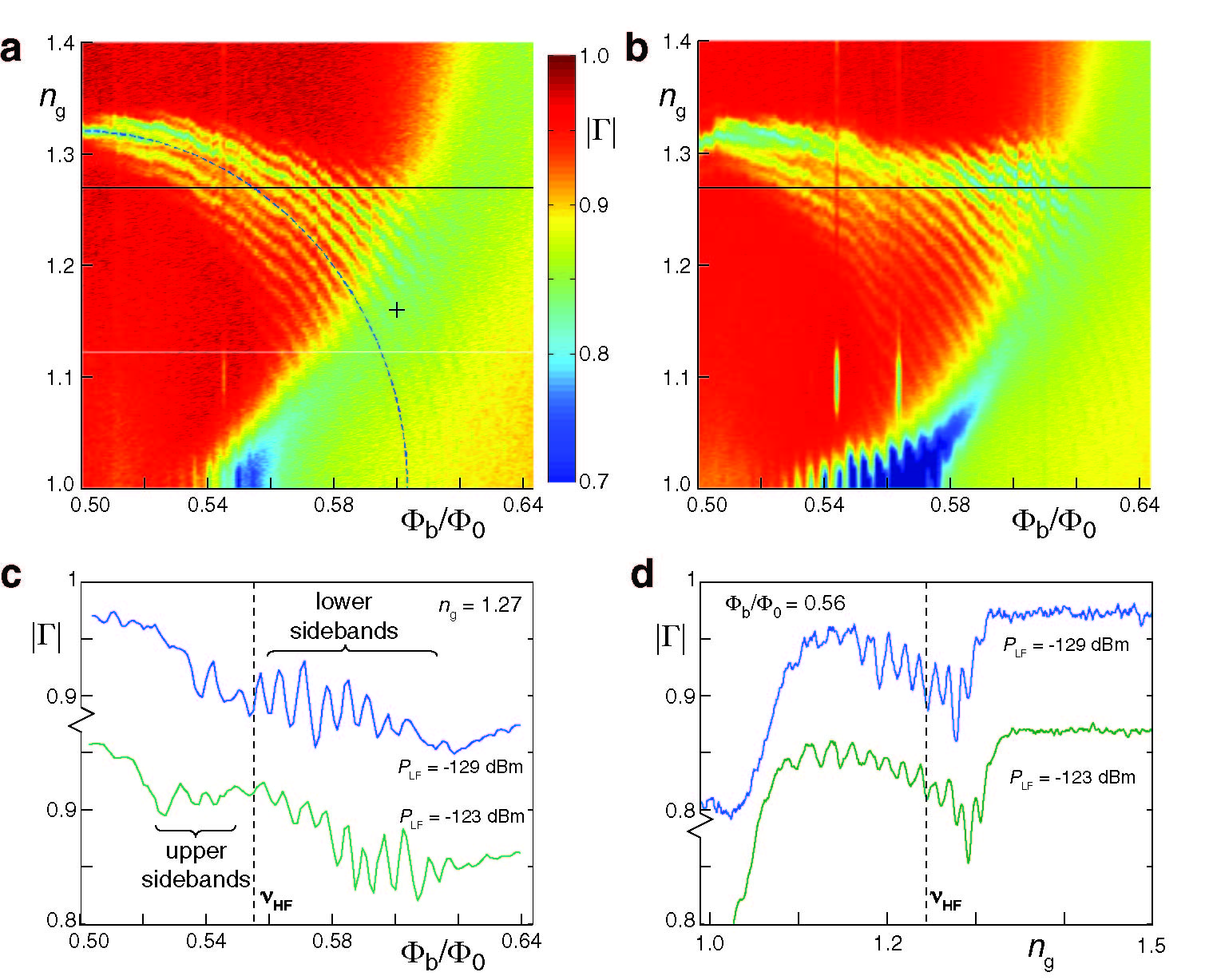}
\caption{Measured spectra of the artificial molecule. 
{\bf a}, The magnitude of the reflection coefficient $|\Gamma|$ as a function of the bias point $(\Phi_b,n_g)$. The color bar gives the scale for  $|\Gamma|$.  
The dashed arch indicates the pure electronic transition
$\Delta E/h=\nu_{\rm HF} =22$ GHz. The cross denotes the bias point of Fig.\ \ref{f.vibronicpot} where $k=-2$. 
The low frequency power $P_{\rm LF} = -129$ dBm. 
{\bf b}, the same as {\bf a} but at higher $P_{\rm LF} = -123$ dBm.
{\bf c},  A cut along $n_g= 1.27$ as indicated by the horizontal lines in panels {\bf a} and {\bf b}. 
{\bf d}, A cut along $\Phi_b= 0.56\Phi_0$.
}
\label{MeasPower}
\end{center}
\end{figure}

In order to study the vibronic transitions, a second drive operating at a high frequency  $\nu_{\rm HF}$ was applied. Figure \ref{MeasPower} shows the results for $\nu_{\rm HF}=22\ $GHz. The reflection coefficient is plotted over the $(\Phi_b,n_g)$ bias plane for two different low-frequency power levels
$P_{\rm LF}$. 
The main observation is the yellowish or light blue fringes. There is one fringe starting at $(\Phi_b,n_g)=(\Phi_0/2, 1.32)$. With increasing $\Phi_b$Ê more fringes appear as they simultaneously are  curving down. One sees that the fringes form concentric and equally spaced ellipses around the point $(\Phi_0/2,1)$. We associate the fringe denoted by the dashed line with a pure electronic transition, and the other fringes with vibronic transitions.  The concentric structure can be understood because in this region the energies are to a good accuracy given by  Eq.\ (\ref{e.Vscoweiow}) and, moreover, $\cos(\pi\Phi/\Phi_0)\approx -\pi(\Phi/\Phi_0-\frac12)$. This means that the difference in the qubit energies increases linearly with the distance in the vicinity of  the degeneracy point $(\Phi_0/2,1)$. In other words, the qubit bands have a conical shape, and the fringes correspond to approximately constant energies. The radii of the fringes are in agreement with the imposed frequencies and the parameters of the qubit, 
$\nu_{\rm HF}=(E^{\rm q}_+-E^{\rm q}_-)/h+k\nu_0$ with integer $k$.  We call the vibronic transitions inside of the dashed line as upper sidebands and outside as lower sidebands, as they are relative to the pure electronic transition at the same bias point ($k>0$ and $k<0$, respectively). 

The visibility  of the fringes is consistent  with the Franck-Condon principle illustrated in Fig.\ \ref{f.vibronicpot}.
In particular, at a half flux quantum ($\Phi_b=\Phi_0/2$) only the pure electronic transition is visible.
This is to be expected since the minima of the potentials $U_\pm$ occur at the same $\Phi$ and thus only diagonal matrix  elements in the vibrational states are nonzero.  For increasing flux the potential minima get shifted, which allows vibronic transitions. The fringes appear in a crescent-shaped region, 
i.e.\ in a stripe that is centered at the basic fringe (dashed line) and whose width grows with $\Phi_b$
(together with the shift of the minima of $U_-$ and $U_+$). It is just in this region
that the classical trajectories of the oscillators intersect and lead to non-vanishing transition matrix elements. The boundary of this region corresponds to trajectories touching each
other at coincident extrema $\Phi_+^{\rm ext}=\Phi_-^{\rm ext}$ in the two electronic states.
Equivalently, for the bias points at the boundary, the oscillation just reaches
the locus of pure electronic transitions (dashed line in Fig.\ \ref{MeasPower}). This allows
to determine the oscillation amplitude corresponding to a given drive power
$P_{\rm LF}$.

For further analysis of the fringes we need to look more closely at the measurement details.
In contrast to experiments with real molecules, we do not detect directly the absorption or emission of the vibronic quanta.
Instead, we measure the reflection of the low-frequency wave. Two effects on the signal can be separated.
1) The first one is a usual resonance absorption taking place when the wave frequency is close to the vibrational frequency $\nu_\pm$. This is different in the upper and lower qubit states but the measurements are consistent with a "motional narrowing" picture that there is only one Lorentzian line whose resonance frequency is an average of $\nu_-$ and $\nu_+$ weighted by the occupations of the qubit states. This is essential for the visibility of the fringes near $\Phi_b\approx\Phi_0/2$  as both the upper and lower state frequencies would be far away to give any noticeable absorption. 
2) The second effect is a damping or amplification caused by the high-frequency drive\cite{Martin04,Grajcar07}.
Consider the lower sideband fringes in Fig.\ \ref{MeasPower}. The absorption of a high-frequency quantum brings the system to a lower vibrational state. When the electronic part of the excitation relaxes by another route, this leads to damping of the vibrations. In a steady state this leads to increased absorption as the low-frequency drive has to supply the missing vibrational quanta. Conversely, in the  case of upper sideband 
the absorption of a high-frequency quantum produces extra vibrational quanta leading to amplification of the oscillation. This could lead to a low-frequency reflection coefficient exceeding unity and even to lasing\cite{Astafiev07}. In our system system the losses in the vibrations cause that we see this only as a decrease in the absorption.

\begin{figure}
\begin{center}
\includegraphics[width=1.0\linewidth]{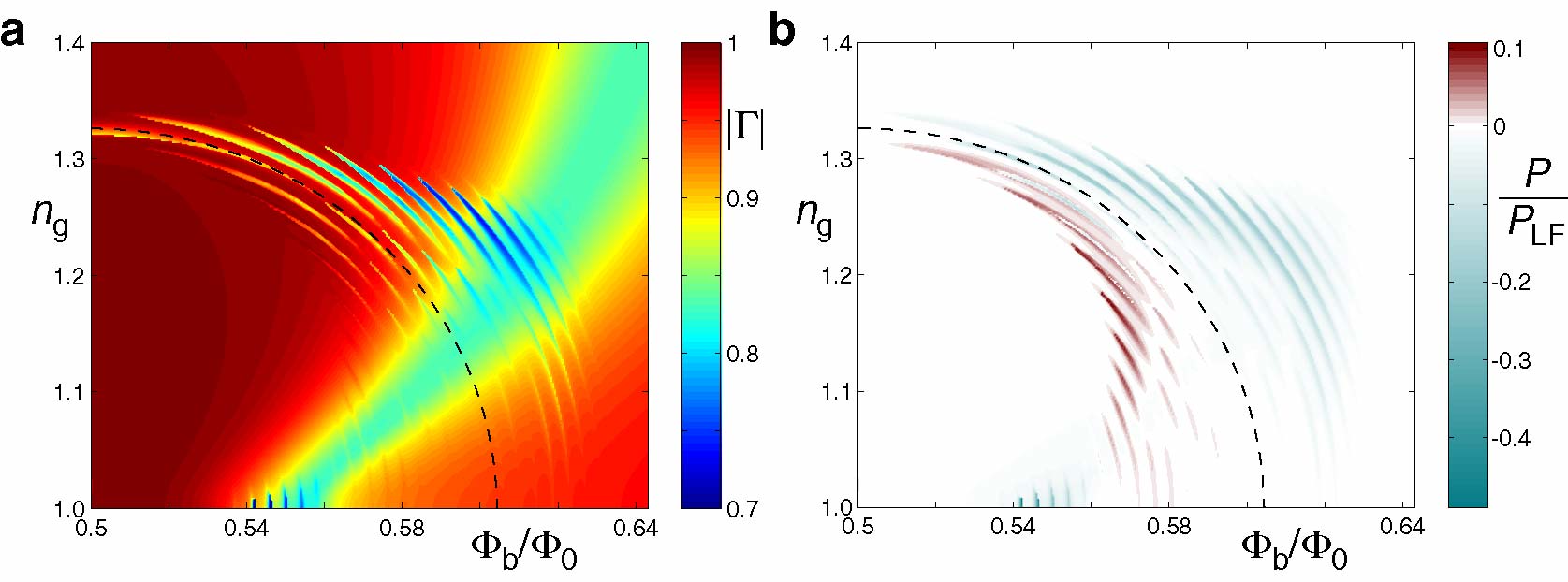}
\caption{Simulated properties  of the artificial molecule. 
{\bf a}, The reflection amplitude $|\Gamma|$ calculated as a function of the
bias  point in the flux-charge plane. The parameters correspond to the measurement plotted in Fig.\ \ref{MeasPower}a. {\bf b}, The energy flow from the qubit to the resonator showing damping for lower sideband transitions (blue)  and amplification for the upper sideband transitions (red). The power unit is $P_{\rm LF}=-129$ dBm.
} \label{simu_gamma}
\end{center}
\end{figure}

The damping/amplification effect of the high-frequency drive provides a natural explanation for the asymmetry of the upper and lower sideband fringes. In order to study this quantitatively, we have made numerical simulations using  Bloch equations for the qubit and coupled to classical equations of the circuit.   The simulated reflection amplitude $|\Gamma|$ is shown in Fig.\ \ref{simu_gamma}a.   Fig.\ \ref{simu_gamma}b shows the energy flow from the qubit to the oscillator. This is defined as $P_{\rm LF,R}-P_{\rm LF,A}$, the low frequency dissipation in the oscillator resistance minus the absorption from the LF drive in the steady state.  We see that this is negative for lower sidebands and positive for upper sidebands, in accordance with the expectation.
The damping on the lower sidebands implies cooling, i.e. the microwaves lead to cooling of the oscillator, similar to side-band cooling in optics\cite{Martin04}. Thus we  see the sideband cooling effect of the vibronic transitions. With a different  measurement method, not driving the object to be cooled, one should get  damping of thermal fluctuations. 

We comment on some additional features of the spectra in Fig.\ \ref{MeasPower}. 
There is a broad feature starting around $(0.54\Phi_0,1)$ and shifting to larger $\Phi_b$ with increasing $n_g$. This is the line where the vibrations in the lower electronic state are in resonance with the low frequency excitation, $\nu_-=\nu_{\rm LF}$. This resonance is present also in the absence of microwave excitation and was used to fit the parameters of the qubit. On the side of smaller $\Phi_b$ the vibronic transitions are well visible as increased absorption because of motional narrowing since $\nu_-<\nu_{\rm LF}<\nu_+$. On the opposite side $\nu_-$ is shifted above $\nu_{\rm LF}$ and the fringes are less clear. There are resonances also around $(0.55\Phi_0,1)$. They also occur in the absence of high-frequency drive. They are caused by Landau-Zener tunneling at the point $(\Phi_0/2,1)$, where the electronic energy difference is at minimum\cite{Oliver05,Sillanpaa06}.




The data were taken at the base temperature of the refrigerator $T=25$ mK.
The high frequency power $P_{\rm HF} = +15$ dBm is measured at the microwave source.
Fig.\ \ref{MeasPower} shows the absolute value of the reflection coefficient defined by $V_{\rm out}=\Gamma V_{\rm in}$. In addition to several stages of
band pass filtering by Minicircuits VLX-series filters, we used two
circulators at 25~mK to prevent the back-action noise of our
cryogenic low-noise amplifier from reaching the qubit. The
attenuation of the rf-signal lines were measured at room
temperature. Cooling down to 25 mK reduced the attenuation by 6 dB.
We checked that the AC Stark shift at large drive powers agreed with
the excitation amplitudes on the measurement drive.

The simulation uses the circuit diagram of Fig.\  \ref{LSET_circ} with $T_1=40$ ns, $T_2=2$ ns and other parameters mentioned in the text. In producing Fig.\ \ref{simu_gamma} the cone approximation mentioned in the text is used as the difference to full Mathieu  bands is negligible.


Fruitful discussions with T.~Heikkil\"a,
F.~Hekking, T. Lehtinen, M.~Paalanen, and M. Sillanp\"a\"a are gratefully
acknowledged. This work was financially supported by the Academy of
Finland, the National Technology Agency, the Finnish Cultural Foundation, the Magnus Ehrnrooth Foundation, EU-INTAS 05-10000008-7923 and
EC-funded ULTI
Project (Contract
RITA-CT-2003-505313).


\end{document}